\documentclass[11pt]{article}
\usepackage{amsfonts}
\usepackage{amsmath}
\usepackage{amssymb}
\usepackage{authblk}
\usepackage{bm}
\usepackage{booktabs}
\usepackage[margin=1in]{geometry}
\usepackage{graphicx}
\usepackage{listings}
\usepackage{comment}
\usepackage{xcolor}
\usepackage{float}
\usepackage[colorlinks=True]{hyperref}
\usepackage{pifont}

\newcommand{\cmark}{\ding{51}}%
\newcommand{\xmark}{\ding{55}}%

\graphicspath{{figs/}}

\title{Systems Biology: Identifiability analysis and parameter identification via systems-biology informed neural networks}
\author[1]{Mitchell Daneker}
\author[2]{Zhen Zhang}
\author[2,3]{George Em Karniadakis}
\author[1,*]{Lu Lu}
\affil[1]{Department of Chemical and Biomolecular Engineering, University of Pennsylvania, Philadelphia, PA, USA}
\affil[2]{Division of Applied Mathematics, Brown University, Providence, RI, USA}
\affil[3]{School of Engineering, Brown University, Providence, RI, USA}
\affil[*]{Corresponding author. Email: lulu1@seas.upenn.edu}
\date{}

\begin{document}

\maketitle

\begin{abstract}
The dynamics of systems biological processes are usually modeled by a system of ordinary differential equations (ODEs) with many unknown parameters that need to be inferred from noisy and sparse measurements. Here, we introduce systems-biology informed neural networks for parameter estimation by incorporating the system of ODEs into the neural networks. To complete the workflow of system identification, we also describe structural and practical identifiability analysis to analyze the identifiability of parameters. We use the ultridian endocrine model for glucose-insulin interaction as the example to demonstrate all these methods and their implementation.
\end{abstract}

\noindent
\textbf{Key words}: Systems biology, Parameter estimation, Structural identifiability, Practical identifiability, Physics-informed neural networks

\section{Introduction}

Systems biology aims to understand biological systems at a system level, including their structures and their dynamics \cite{kitano2002systems}. Often, a biological system is modeled by a system of ordinary differential equations (ODEs), which describes the dynamics of the various concentrations of chemical and molecular species as a function of time. These biological models usually introduce some parameters that are unknown and required to be estimated accurately and efficiently. Hence, one central challenge in systems biology is the estimation of unknown model parameters (e.g., rate constants), after which we can perform the prediction of model dynamics. Parameter estimation requires observations of the state variables of the system, but due to technical limitations, only part of the state variables are observable in experiments, which makes parameter estimation even more difficult.

\begin{figure}[htbp]
    \centering
    \includegraphics[width=\textwidth]{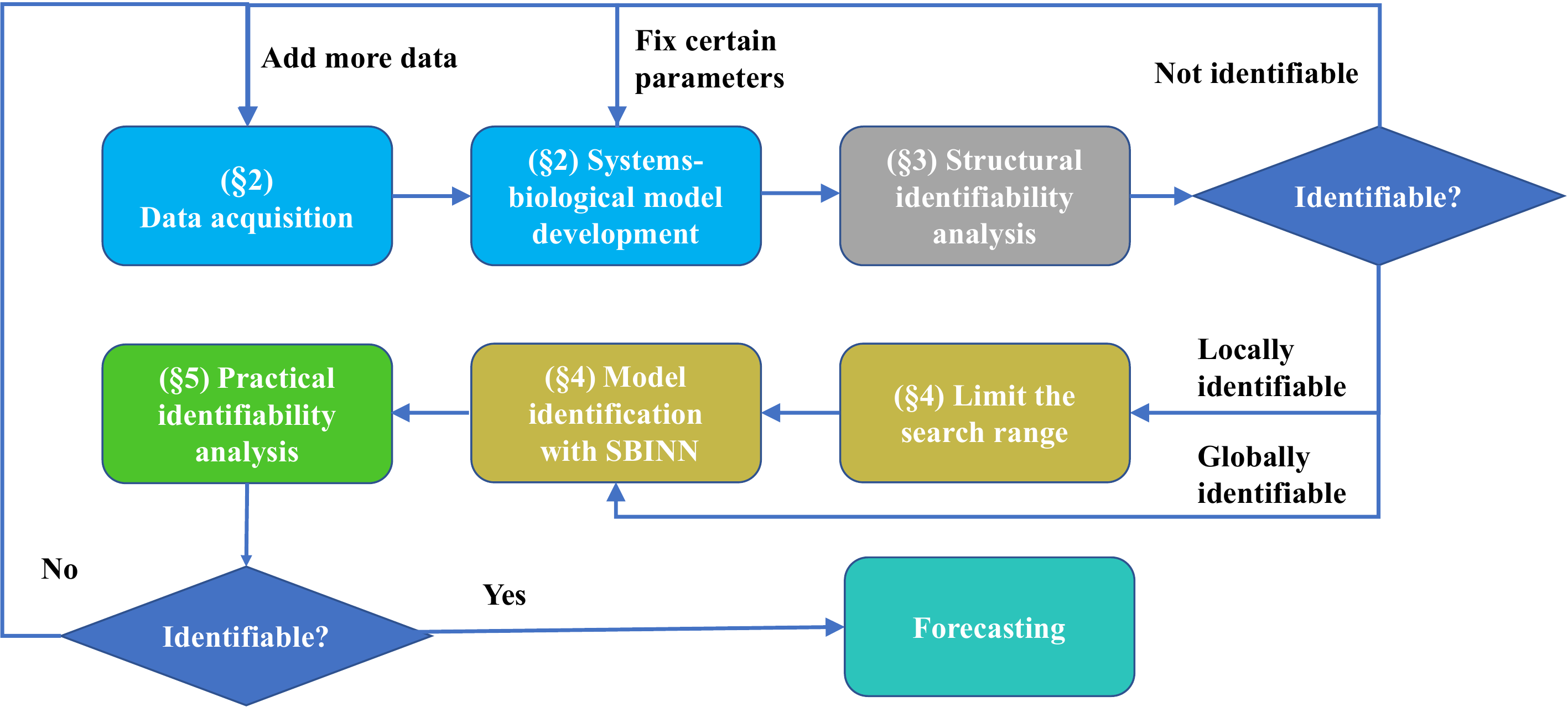}
    \caption{\textbf{The workflow for the development and identification of a systems biological model.}}
    \label{fig:flowchart}
\end{figure}

In this chapter, we introduce the workflow for the development and identification of systems biological models (Fig.~\ref{fig:flowchart}). The whole workflow includes the following several steps:
\begin{itemize}
    \item Step 1: Data acquisition and systems-biological model development (Section~\ref{sec:model-definition}). As the first step, we need to collect experimental data for the underlying system and develop ODEs to model the system dynamics. This is not the focus of this chapter, and we directly use the ultradian endocrine model for glucose-insulin interaction \cite{sturis1991computer}.
    \item Step 2: Structural identifiability analysis (Section~\ref{sec:struc-ident}). With a proposed model, we determine which parameters of the model are structurally identifiable. If the parameters are not structurally identifiable, Step 1 is revisited such as adding more data or fixing certain parameters. If the parameters are locally identifiable, we need to limit their search range.
    \item Step 3: Parameter estimation via systems-biology informed neural network (SBINN) (Section~\ref{sec:SBINN}). We next use a SBINN to infer the unknown model parameters from the data.
    \item Step 4:  Practical identifiability analysis (Section~\ref{sec:pract-ident}). With the inferred parameters, we check the quality of the estimates via practical identifiable analysis. If the parameters are practically identifiable, we can use the identified model for forecasting, otherwise, we need to revisit Step~1.
\end{itemize}
The code used in the chapter is publicly available from the GitHub repository \url{https://github.com/lu-group/sbinn}.

\section{Ultradian endocrine model for glucose-insulin interaction}
\label{sec:model-definition}

To demonstrate the methods, we consider the system of the glucose-insulin interactions and use a relatively simple ultradian model \cite{sturis1991computer} with 6 state variables and 21 parameters. The state variables are plasma insulin concentration $I_{p}$, interstitial insulin concentration $I_{i}$, glucose concentration $G$, and a three stage filter $(h_1,h_2,h_3)$ that mimics the response of the plasma insulin to glucose levels.

Eqs.~\eqref{eq:glucose1} and \eqref{eq:glucose2} provide the system of equations for the model where major parameters include (i) $E$, a rate constant for exchange of insulin between the plasma and remote compartments; (ii) $I_G$, the exogenous (externally driven) glucose delivery rate; (iii) $t_p$, the time constant for plasma insulin degradation; (iv) $t_i$, the time constant for the remote insulin degradation; (v) $t_d$, the delay time between plasma insulin and glucose production; (vi) $V_p$, the volume of insulin distribution in the plasma; (vii) $V_i$, the volume of the remote insulin compartment; (viii) $V_g$, the volume of the glucose space \cite{sturis1991computer,albers2017personalized}. Furthermore, in Eq.~\eqref{eq:glucose2}, $f_1(G)$ provides the rate of insulin production; $f_2(G)$ defines insulin-independent glucose utilization; $f_3(I_i)$ is the insulin-dependent glucose utilization; and $f_4(h_3)$ represents delayed insulin-dependent glucose utilization.

\begin{subequations}\label{eq:glucose1}
\begin{eqnarray}
\frac{dI_p}{dt} =  f_1(G)-E\bigl(\frac{I_{p}}{V_{p}}-\frac{I_i}{V_{i}}\bigr)-\frac{I_{p}}{t_{p}}, && \frac{dI_i}{dt} = E\bigl(\frac{I_{p}}{V_{p}}-\frac{I_i}{V_{i}}\bigr)-\frac{I_{i}}{t_{i}},\\
\frac{dG}{dt} = f_4(h_3)+I_{G}(t)-f_2(G)-f_3(I_i)G, && \frac{dh_1}{dt} = \frac{1}{t_d}\bigl(I_p-h_1\bigr), \\
\frac{dh_2}{dt} = \frac{1}{t_d}\bigl(h_1-h_2\bigr), && \frac{dh_3}{dt} = \frac{1}{t_d}\bigl(h_2-h_3\bigr),
\end{eqnarray}
\end{subequations}
where $f_1$--$f_4$ and the nutritional driver of the model $I_G(t)$ are given by
\begin{subequations}\label{eq:glucose2}
\begin{eqnarray}
f_1(G) = \frac{R_m}{1+ \exp(\frac{-G}{V_g c_1} + a_1)}, && f_2(G) = U_b \left(1-\exp(\frac{-G}{C_2V_g}) \right), \\
f_3(I_i) = \frac{1}{C_3 V_g} \left( U_0 + \frac{U_m}{1+(\kappa I_i)^{-\beta}} \right), && f_4(h_3) = \frac{R_g}{1 + \exp(\alpha (\frac{h_3}{C_5 V_p}-1))}, \\
\kappa = \frac{1}{C_4} \left(\frac{1}{V_i} + \frac{1}{E t_i} \right), && I_G(t) = \sum^N_{j=1}{m_j k\exp(k(t_j-t))},
\end{eqnarray}
\end{subequations}
where the nutritional driver $I_G(t)$ is a systematic forcing term that acts as nutritional intake of glucose and is defined over $N$ discrete nutrition events \cite{albers2014dynamical} with $k$ as the decay constant and event $j$ occurs at time $t_j$ with carbohydrate quantity $m_j$. The nominal values of the parameters are provided in Table \ref{table:glucose-app}.

\begin{table}[htbp]
\centering
\caption{\textbf{Parameters for the ultradian glucose-insulin model \cite{albers2017personalized}.} The search range of the first 7 parameters is adopted from \cite{sturis1991computer}, and the search range of the other parameters is $(0.2p^*, 1.8p^*)$, where $p^*$ is the nominal value of that parameter.}
\begin{tabular}{ccccc}
\toprule
Parameter & Nominal value & Unit & Search range & Inferred Value \\
\midrule
$V_p$  & $3$ & $lit$ & -- & --\\ \hline
$V_i$  & $11$ & $lit$ & -- & --\\ \hline
$V_g$ & $10$ & $lit$ & -- & --\\ \hline
$E$  & $0.2$ & $lit \ min^{-1}$ & (0.100, 0.300) &  $0.201$\\ \hline
$t_p$  & $6$ & $min$ & (4.00, 8.00) & $5.99$ \\ \hline
$t_i$  & $100$ & $min$ & (60.0, 140) & $101.20$ \\ \hline
$t_d$  & $12$ & $min$ & (25/3, 50/3) & $11.98$ \\ \hline
$k$  & $0.0083$ & $min^{-1}$ & (0.00166, 0.0150) & $0.00833$ \\ \hline
$R_m$  & $209$ & $mU \ min^{-1}$ & (41.8, 376) & $208.62$ \\ \hline
$a_1$  & $6.6$ & & (1.32, 11.9) & $6.59$ \\ \hline
$C_1$  & $300$ & $mg \ lit^{-1}$ & (60.0, 540) & $301.26$ \\ \hline
$C_2$  & $144$ & $mg \ lit^{-1}$ & (28.8, 259) & $37.65$ \\ \hline
$C_3$  & $100$ & $mg \ lit^{-1}$ & -- & -- \\ \hline
$C_4$  & $80$ & $mU \ lit^{-1}$ & (16.0, 144) & $78.76$ \\ \hline
$C_5$  & $26$ & $mU \ lit^{-1}$ & (5.20, 46.8) & $25.94$ \\ \hline
$U_b$  & $72$ & $mg \ min^{-1}$ & (14.4, 130) & $71.33$ \\ \hline
$U_0$  & $4$ & $mg \ min^{-1}$ & (0.800, 7.20) & $0.0406C_3$ \\ \hline
$U_m$  & $90$ & $mg \ min^{-1}$ & (18.0, 162) & $0.890C_3$ \\ \hline
$R_g$  & $180$ & $mg \ min^{-1}$ & (36.0, 324) & $179.86$ \\ \hline
$\alpha$  & $7.5$ & & (1.50, 13.5) & $7.54$ \\ \hline
$\beta$  & $1.772$ & & (0.354, 3.190) & $1.783$ \\
\bottomrule
\end{tabular}
\label{table:glucose-app}
\end{table}

\begin{figure}[htbp]
    \centering
    \includegraphics{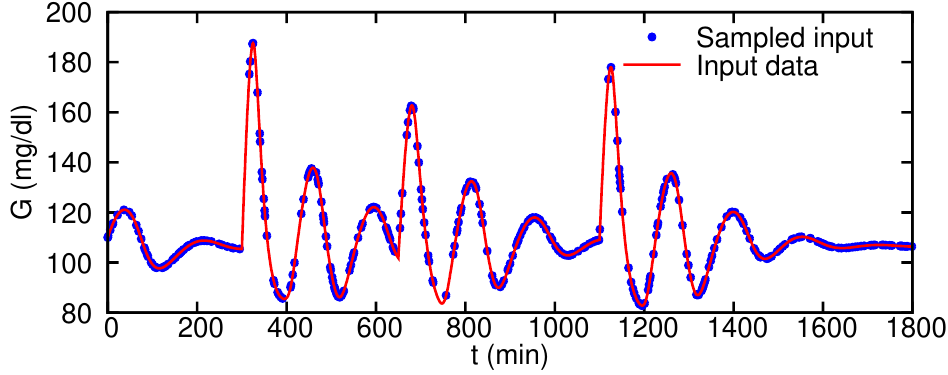}
    \caption{\textbf{Ultradian glucose-insulin model observation data for parameter inference.} 360 measurements on glucose level ($G$) only are randomly sampled in the time window of $0-1800$ minutes ($\sim$ one day). Figure is adapted with permission from \cite{yazdani2020systems}.}
    \label{fig:glucose-input}
\end{figure}

Synthetic data is generated by numerically solving the system from time $t=0$ to $t=1800 \ min$ with the initial conditions $\mathbf{x}(0)=[12.0\ (\mu U/ml) \ 4.0\ (\mu U/ml) \ 110.0\ (mg/dl) \ 0.0 \ 0.0 \ 0.0]$ and three nutrition events $(t_j, m_j) = [(300, 60) \ (650, 40) \ (1100, 50)] \ (min, g)$ pairs. This completes the first two steps of the flowchart in Fig~\ref{fig:flowchart}. We assume the only observable is the glucose level measurements $G$, which are sampled randomly, as shown in Fig. \ref{fig:glucose-input}.

\section{Structural identifiability analysis}
\label{sec:struc-ident}

In this section, we investigate whether a set of unknown parameters in the Ultradian endocrine model for glucose-insulin interaction is structurally identifiable from the glucose concentration data $G$. Simply fitting a model to the data is not sufficient to show how reliable the estimated parameters are. Insufficient data can produce very different sets of parameters without affecting the fit of data if a model is structurally non-identifiable. To resolve the non-identifiability issue, there are two options: one is to acquire data for more species, another is to fix certain parameters as their nominal values.

Suppose we are given a dynamical system of the following abstract form
\begin{equation*}
    X' = f(X,\Theta, u), \quad y = g(X,\Theta, u),
\end{equation*}
where $X = (X_1,\cdots,X_n)$ represents the state variables, $y= (y_1,\cdots,y_m)$ represents the observables. $\Theta = \left(\theta_1,\cdots,\theta_k\right)$ contains the parameters to identify, and $u$ represents the input variable to the system. A parameter set $\Theta$ is called structurally \textit{globally} identifiable if
\begin{equation}\label{eq:struc_model}
    g(X,\Theta, u) = g(X,\Phi, u) \quad\implies \quad\Theta = \Phi
\end{equation}
for every $\Phi = (\phi_1,\cdots,\phi_k)$ in the same space as $\Theta$.
\textit{Local} identifiability only requires Eq.~(\ref{eq:struc_model}) to hold in a neighbourhood of $\Theta$. As a consequence, if a model parameter turns out to be locally identifiable, it is suggested that one should limit the search range for this parameter before fitting the model. For globally identifiable parameters, this step is not required.

In this section, we only test for the local identifiability of the system since the existing software packages may suffer from out-of-memory issues when testing the global identifiability of a system with a large number of state variables and a small number of observables. For convenience, we will refer to a system as being identifiable when it is structually locally identifiable. We use the Julia library \textit{StructuralIdentifiability} \cite{structidjl} to test for structural identifiability of the model. The existing algorithms implemented in the library require both $f$ and $g$ to be rational functions, which are fractions of polynomials.

\subsection{Preprocessing}

There are exponential functions and power functions in Eq.~\eqref{eq:glucose2} of the model, and thus a preprocessing step is required to get rid of the transcendental components of the system. One method is to introduce a set of extra state variables $g_i$ which are equal to these transcendental components and apply the chain rule to find their derivatives. In our example, one can set
\begin{subequations}
\begin{eqnarray*}
   g_1(t) = 1+ \exp(\frac{-G(t)}{V_g C_1} + a_1), &&
   g_2(t) = 1-\exp(\frac{-G(t)}{C_2V_g}),\\
   g_3(t) = 1+(\kappa I_i(t))^{-\beta}, &&
   g_4(t) = 1 + \exp(\alpha (\frac{h_3(t)}{C_5 V_p}-1)).
\end{eqnarray*}
\end{subequations}
It follows from the chain rule that 
\begin{subequations}
\begin{eqnarray*}
   \frac{dg_1}{dt} = -\frac{g_1 -1}{V_gC_1}\frac{dG}{dt}, &&
   \frac{dg_2}{dt} = -\frac{g_2 -1}{V_gC_1}\frac{dG}{dt},\\
   \frac{dg_3}{dt} = -\beta k\frac{g_3-1}{kI_i}\frac{dI_i}{dt}, &&
   \frac{dg_4}{dt} = \frac{\alpha}{C_5V_p}(g_4-1)\frac{dh_3}{dt}.
\end{eqnarray*}
\end{subequations}

The ODE system in Eqs.~\eqref{eq:glucose1} and \eqref{eq:glucose2} can be rewritten in the following rational form:
\begin{subequations}\label{eq:new_system}
\begin{eqnarray}
	\frac{dI_p}{dt} = \frac{R_m}{g_1} - E(\frac{I_p}{V_p}-\frac{I_i}{V_i})-\frac{I_p}{t_p}, && \frac{dI_i}{dt} = E(\frac{I_p}{V_p}-\frac{I_i}{V_i})-\frac{I_i}{t_i},\\
	\frac{dG}{dt} = \frac{R_g}{g_4} + I_G - U_bg_2-\frac{G}{C_3V_g}(U_0+\frac{U_m}{g_3}), && \frac{dg_1}{dt} =-\frac{g_1-1}{V_gC_1}\frac{dG}{dt},\\
	\frac{dg_2}{dt} =-\frac{g_2-1}{V_gC_2}\left(\frac{R_g}{g_4} + u_1 - U_bg_2-\frac{G}{C_3V_g}(U_0+\frac{U_m}{g_3})\right), && \frac{dg_3}{dt} =-\beta \kappa\frac{g_3-1}{\kappa I_i}\frac{dI_i}{dt},\\
	\frac{dg_4}{dt} = \frac{\alpha}{C_5V_pt_d}(g_4-1)(h_2-h_3), &&
	\frac{dh_1}{dt} =\frac{1}{t_d}(I_p-h_1),\\
	\frac{dh_2}{dt} =\frac{1}{t_d}(h_1-h_2), &&
	\frac{dh_3}{dt} =\frac{1}{t_d}(h_2-h_3),
\end{eqnarray}
\end{subequations}
where $I_G$ is treated as the input to the system and $G$ is the output/observable of the system.
Note that the initial conditions of all the ODE systems are assumed to be unknown for the \textit{StructuralIdentifiability} library to work. This means there are 4 extra degrees of freedom lying in the initial conditions of $g_1,g_2,g_3,g_4$ in Eq.~\eqref{eq:new_system} compared to Eqs.~\eqref{eq:glucose1} and \eqref{eq:glucose2}. Consequently, any identifiable parameter in the new system will be identifiable in the original system, but not the other way around. Our goal now reduces to find a set of identifiable parameters in Eq.~\eqref{eq:new_system}.

\subsection{Structural identifiability results}

\begin{figure}[htbp]
\centering
\includegraphics{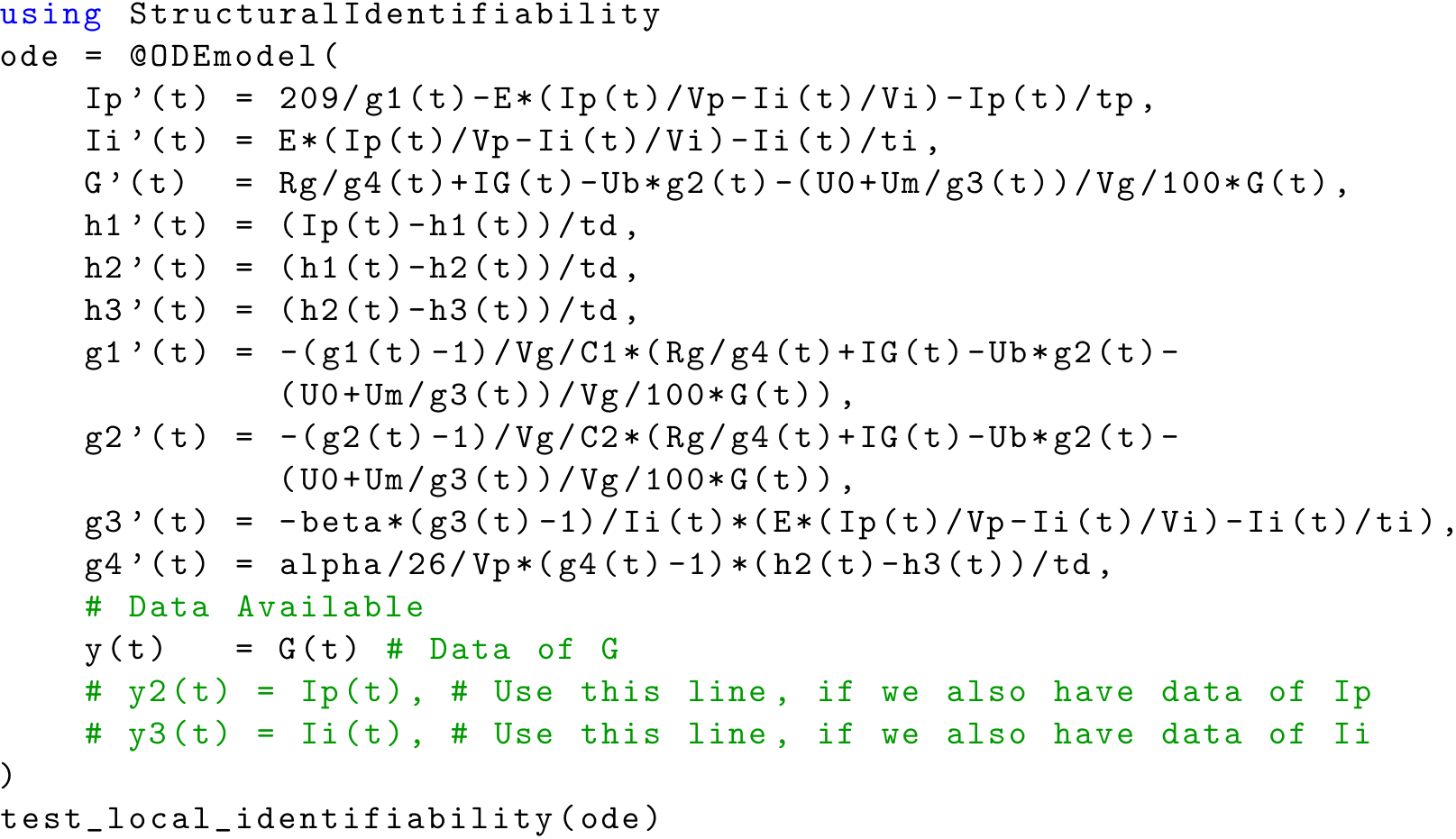}
\caption{\textbf{Specify the ODE model.} We specify the parametric ODE model in Eq.~\eqref{eq:new_system} using the \texttt{@ODEmodel} macro. \texttt{x'(t)} is the derivative of state variable \texttt{x(t)}, which is assumed to be unknown if not specified otherwise. \texttt{y(t)} defines the output variable which is assumed to be given. The last line tests the local identifiability of the model.}
\label{code:julia_model}
\end{figure}

By inspection of the three scaling invariances of the ODE in Eq.~\eqref{eq:new_system}, namely
\begin{equation}\label{eq:invariance}
    \begin{cases}
        R_m = cR_m \\
        I_p(0) = cI_p(0) \\
        I_i(0) = cI_i(0)
    \end{cases},\quad
    \begin{cases}
        \alpha = c\alpha \\
        C_5 = cC_5 
    \end{cases},
    \begin{cases}
        U_0 = cU_0 \\
        U_m = cU_m \\
        C_3 = cC_3
    \end{cases},
\end{equation}
we found intrinsic structural non-identifiability of these parameters of the model. To get rid of the scaling invariances, one needs to fix one parameter in each system of equations of Eq.~\eqref{eq:invariance}. For illustration purposes, we fix $R_m$, $C_3$, and $C_5$ as their nominal values in Table~\ref{table:glucose-app} as an example and check whether the rest of the parameters are locally identifiable, see the code in Fig.~\ref{code:julia_model}.

Here, 15 undetermined parameters are remaining in the modified system, and it is impossible to fit all of them simultaneously, as shown in the first row of Table~\ref{table:struc_moredata}. This is reasonable because we assume that there is only one observable $G$ and it is hard to infer all parameters with limited amount of data. As demonstrated in Fig.~\ref{fig:flowchart}, to resolve the identifiability issue, one possible option is to acquire more data. It can be observed from the second row of Table~\ref{table:struc_moredata} that taking $I_i$ and $I_G$ as additional observables makes $t_p$ and $t_i$ locally identifiable. Still, a large proportion of parameters remain structurally non-identifiable.

The second option (fixing certain parameters) is also considered. Here, we consider three different cases, where $(V_p)$, $(V_p,V_i)$ and $(V_p,V_i,V_g)$ are fixed respectively. We still assume that we only have the glucose concentration $G$ available. In the fourth and fifth rows of Table~\ref{table:struc_moredata}, we see that more parameters become identifiable when we fix $V_p$ and $V_i$, but the model is still not identifiable. It is only identifiable when all three parameters are set as fixed values.


\begin{table}[htbp]
\centering
\caption{\textbf{Local structural identifiability result of the ultradian endocrine model with different observables and parameters.} $R_m$, $C_3$ and $C_5$ are prefixed. More parameters become structurally identifiable when more data ($I_p$ and $I_i$) are given. With only $G$ given, the model is structurally locally identifiable when $V_p, V_i, V_g$ are fixed.}
\label{table:struc_moredata}
\begin{tabular}{cccccccccccccccc}
\hline
Parameter  & $V_p$  & $V_i$  & $V_g$  & $E$   & $t_p$  & $t_i$  & $t_d$  & $C_1$  & $C_2$  & $U_b$ & $U_0$ & $U_m$ & $R_g$ & $\alpha$ & $\beta$ \\ \hline
Given $G$ & \xmark & \xmark & \xmark & \xmark & \xmark & \xmark & \cmark & \xmark & \xmark & \cmark & \xmark & \xmark & \cmark & \xmark & \cmark \\ \hline
Given $G, I_p, I_i$ & \xmark & \xmark & \xmark & \xmark & \cmark & \cmark & \cmark & \xmark & \xmark & \cmark & \xmark & \xmark & \cmark & \xmark & \cmark \\ \hline
Given $G$ & -- & \xmark & \xmark & \xmark & \xmark  & \xmark  & \cmark  & \xmark & \xmark & \cmark & \xmark & \xmark & \cmark  & \cmark & \cmark  \\ \hline
Given $G$ & -- & -- & \xmark & \cmark & \cmark & \cmark & \cmark & \xmark & \xmark & \cmark & \xmark & \xmark & \cmark & \cmark & \cmark \\ \hline
Given $G$ & -- & -- & -- & \cmark & \cmark & \cmark & \cmark & \cmark & \cmark & \cmark & \cmark & \cmark & \cmark & \cmark & \cmark \\ \hline
\end{tabular}
\end{table}

In summary, the ODE system of Eq.~\eqref{eq:new_system} is structurally locally identifiable when $R_m$, $C_3$, $C_5$, $V_p$, $V_i$ and $V_g$ are fixed. As a final step, we relate the identifiability of the modified system to original Ultradian glucose-insulin model described by Eqs.~\eqref{eq:glucose1} and \eqref{eq:glucose2}. Note that the scaling invariance between $R_m$ and $I_p(0), I_i(0)$ breaks when the later two are provided in the training as the initial conditions. Also the scaling invariance between $\alpha$ and $C_5$ does not hold in the original system, since the value for $\alpha$ can be uniquely determined by the initial condition for $g_4$. The scaling invariance for $U_0$, $U_m$ and $C_3$ still holds in Eqs.~\eqref{eq:glucose1} and \eqref{eq:glucose2}, but one can expect $U_0/C_3$, $U_m/C_3$ to be a constant. Therefore, one needs to only to fix $V_p$, $V_i$ and $V_g$ in the parameter estimation process.

\section{Parameter estimation via SBINN}
\label{sec:SBINN}

\subsection{Deep neural networks}
\label{sec:DNN}

\begin{figure}[htbp]
    \centering
    \includegraphics[width=6cm]{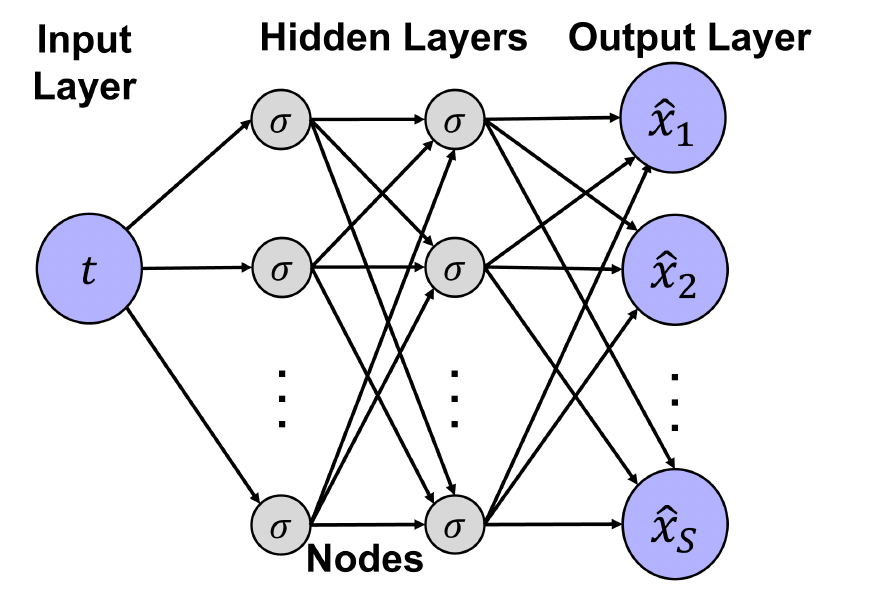}
    \caption{\textbf{Architecture of a fully connected neural network.} A neural network consists of an input layer (the input $t$), several hidden layers (composed of weights $\bm{W}^{\ell}$, bias $\bm{b}^{\ell}$, and activation function $\sigma$), and an output layer.}
    \label{fig:nn}
\end{figure}

Deep neural networks (DNNs) function by recursively transforming inputs linearly and nonlinearly, i.e., compositional functions. Many types of DNNs have been developed such as convolutional neural networks and recurrent neural networks, and here we only consider fully connected neural networks (FNNs). A FNN is composed of many layers (Fig.~\ref{fig:nn}). We denote a $L$-layer neural network (i.e., $(L-1)$ hidden layers) by $\mathcal{N}^L(\mathbf{x}): \mathbb{R}^{d_{\text{in}}} \to \mathbb{R}^{d_{\text{out}}}$, where $d_{\text{in}}$ and $d_{\text{out}}$ are the dimensions of the input and output, respectively. Each layer has a number of neurons, which can be thought as data processors, which take the output of the previous layer as the input, transform it, and then provide the output to the next layer. We use $N_\ell$ to denote the number of neurons in the $\ell$-th layer. At the input layer we have $N_0 = d_{\text{in}}$, and at the output layer we have $N_L = d_{\text{out}}$.

To define a FNN rigorously, in the $\ell$-th layer, we define a weight matrix $\bm{W}^\ell$, a bias $\mathbf{b}^\ell$, and an activation function $\sigma$. Examples of $\sigma$ include logistic sigmoid ($1/(1+e^{-x})$), the hyperbolic tangent ($\tanh$), and the rectified linear unit (ReLU, $\max\{x, 0\}$). Then a FNN is defined as:
\begin{align*}
    \text{input layer:} & \quad \mathcal{N}^0(\textbf{x}) = \textbf{x} \in \mathbb{R}^{d_{\text{in}}}, \\
    \text{hidden layers:} & \quad \mathcal{N}^\ell(\textbf{x}) = \sigma(\bm{W}^{\ell}\mathcal{N}^{\ell-1}(\textbf{x}) + \bm{b}^{\ell}) \in \mathbb{R}^{N_\ell}, \quad \text{for} \quad 1 \le \ell \le L-1, \\
    \text{output layer:} & \quad \mathcal{N}^{L}(\textbf{x}) = \bm{W}^{L}\mathcal{N}^{L-1}(\textbf{x}) + \bm{b}^{L} \in \mathbb{R}^{d_{\text{out}}}.
\end{align*}
All the weights and biases are the neural network parameters $\boldsymbol{\theta}$.

\subsection{Systems-biology informed neural networks (SBINN)}

SBINN was proposed in \cite{yazdani2020systems} and uses systems-biological models (e.g., Eqs.~\eqref{eq:glucose1} and \eqref{eq:glucose2}) to inform a deep neual network. The network input is time $t$, and the output is a vector of state variables $\hat{\mathbf{x}}(t; \boldsymbol{\theta}) = (\hat{x}_1(t; \boldsymbol{\theta}), \hat{x}_2(t; \boldsymbol{\theta}), \dots, \hat{x}_S(t; \boldsymbol{\theta}))$, which acts as a proxy to the ODE solution.

\begin{figure}[htbp]
\centering
\includegraphics{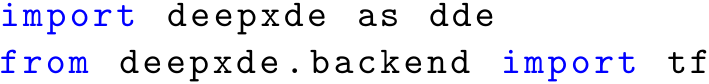}
\caption{\textbf{Importing DeepXDE and the TensorFlow backend.}}
\label{code:python_import}
\end{figure}

We use Python to implement the code, see Appendix \ref{AppendixA} for an introduction to Python. We can directly implement SBINN using general deep learning frameworks such as TensorFlow \cite{abadi2016tensorflow} and PyTorch \cite{paszke2019pytorch}, but the implementation becomes much easier if we use the open-source library DeepXDE \cite{lu2019deepxde}. DeepXDE is a library for scientific machine learning and can use either TensorFlow or PyTorch as its computational engine (called backend). We begin with importing DeepXDE and the backend being used (Fig.~\ref{code:python_import}). Here we choose TensorFlow as the backend, and the code for PyTorch backend is almost the same.

\begin{figure}[htbp]
\centering
\includegraphics{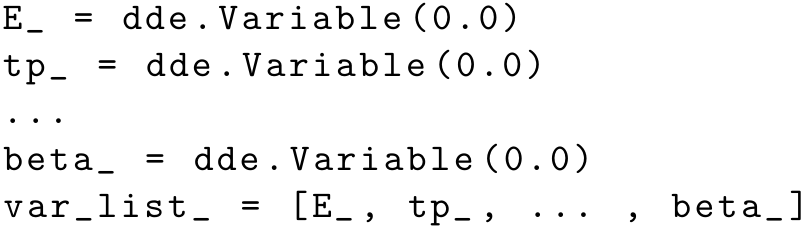}
\caption{\textbf{Creating parameters to estimate.} We initialize all parameters to zero and create a list of these parameters.}
\label{code:sbinn}
\end{figure}

We then implement SBINN. As the first step, we define all parameters to estimate (all the parameters in Table \ref{table:glucose-app} except $V_p$, $V_i$ and $V_g$, which are easily measurable) with an initial guess of zero using \verb|dde.Variable|, and create a list of all the variables to be used later (Fig.~\ref{code:sbinn}).

\begin{figure}[htbp]
\centering
\includegraphics{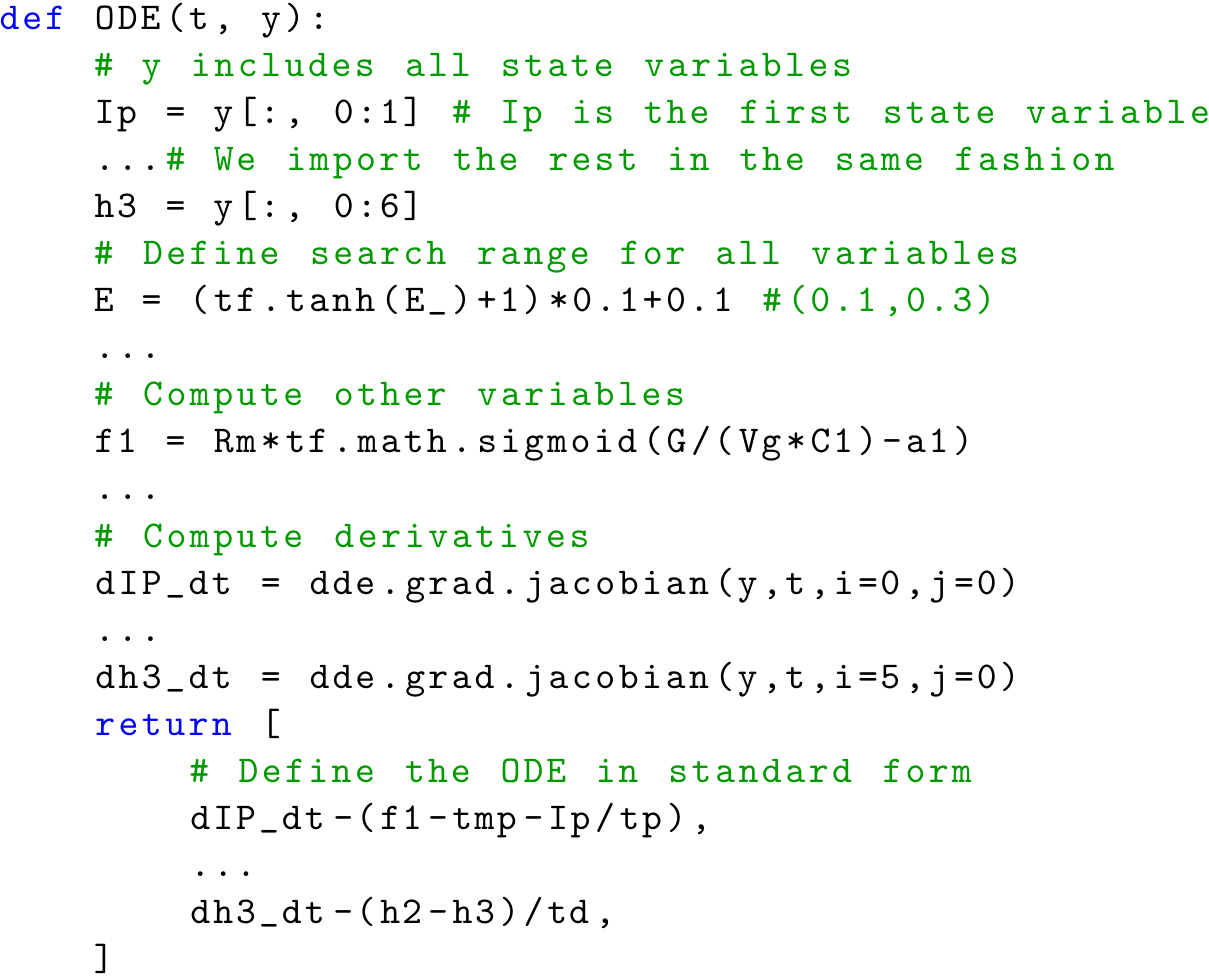}
\caption{\textbf{Implementation of the ODE system in Eqs.~\eqref{eq:glucose1} and \eqref{eq:glucose2}.} The Python function \texttt{ODE} returns the residuals of all ODEs, i.e., the difference between the left-hand size and the right-hand size of the ODE.}
\label{code:ode}
\end{figure}

Next we use these parameters to implement the ODEs for the system. Because we only use the observation of $G$, based on our structural identifiability analysis, we need to limit the search range for the parameters. In this case, the range of seven parameters is adopted from \cite{sturis1991computer}, and the range for other parameters is set as $(0.2p^*, 1.8p^*)$, where $p^*$ is the nominal value of that parameter (Table~\ref{table:glucose-app}). We implement the search range and the ODE system of Eqs.~\eqref{eq:glucose1} and \eqref{eq:glucose2} in Fig.~\ref{code:ode}.

\begin{figure}[htbp]
\centering
\includegraphics{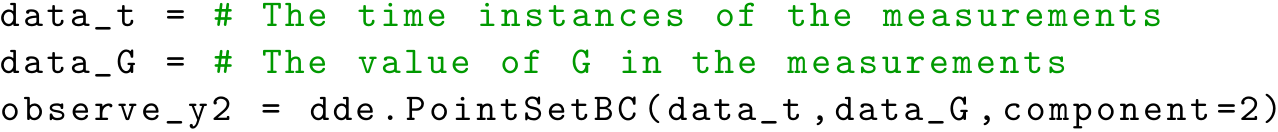}
\caption{\textbf{Implementation of the data observation of $G$.}}
\label{code:BC}
\end{figure}

The next step is to import the data measurements of glucose concentration $G$ via \texttt{dde.PointSetBC} (Fig.~\ref{code:BC}). Other data measurements such as the initial conditions can be done similarly.

\begin{figure}[htbp]
    \centering
    \includegraphics{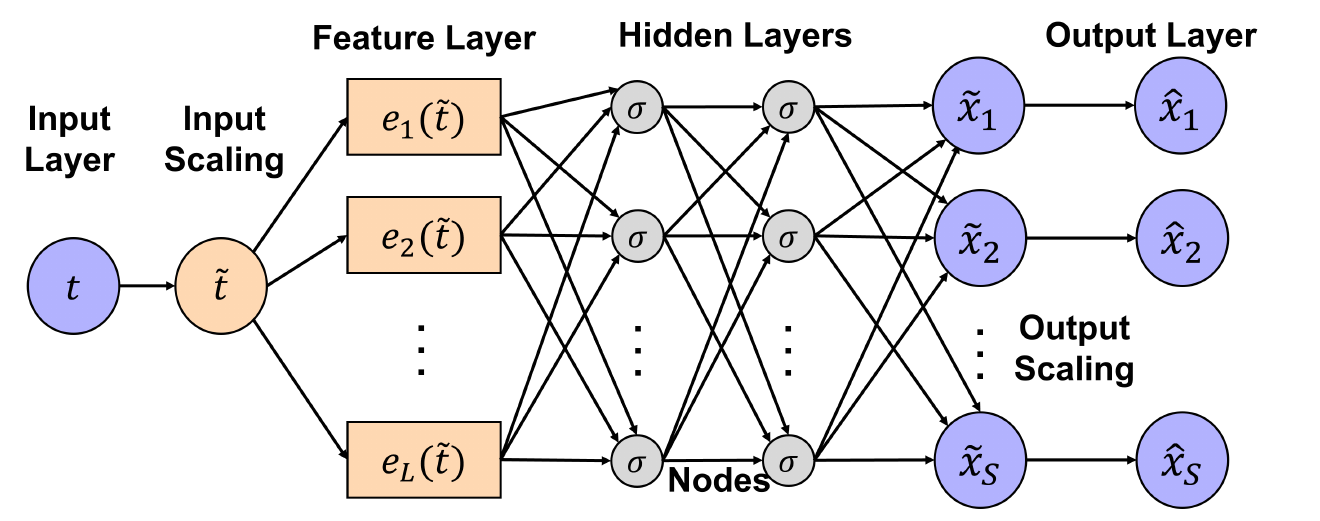}
    \caption{\textbf{Neural network architecture for SBINN.} The input-scaling layer and output-scaling layer scale the network input and outputs to order one. The feature layer provides features directly to the first fully-connected layer.}
    \label{fig:featurenn}
\end{figure}

We have implemented the ODE system and data measurements. Next, we build our neural network model. In order to speed up network training, rather than just the FNN described in Section~\ref{sec:DNN}, we can add additional layers described as follows (Fig.~\ref{fig:featurenn}).
\begin{itemize}
\item Input-scaling layer. In the case of a large time domain, $t$ varies by multiple orders of magnitude, which negatively effects our NN training. We apply a linear scaling function to $t$ using the maximum value of the time domain $T$ to create $\tilde{t} = t/T$ which will be $\sim \mathcal{O}(1)$. 
\item Feature layer. Often, ODE solutions have a pattern such as periodicity or exponential decay. Rather than letting a NN determine these features on its own, we add these patters in a feature layer. Though feature choice is problem specific, the setup is similar for any problem. We use the $L$ function $e_1(\cdot), e_2(\cdot), \dots, e_L(\cdot)$ to construct the $L$ features $e_1(\tilde{t}), e_2(\tilde{t}), \dots, e_L(\tilde{t})$ as seen in Fig~\ref{code:feature_transform}. If no pattern is easily identifiable, it is best to leave out the feature layer than to include something incorrect; this is just a technique to aid in training, not a requirement for the SBINN to work. 

\begin{figure}[htbp]
\centering
\includegraphics{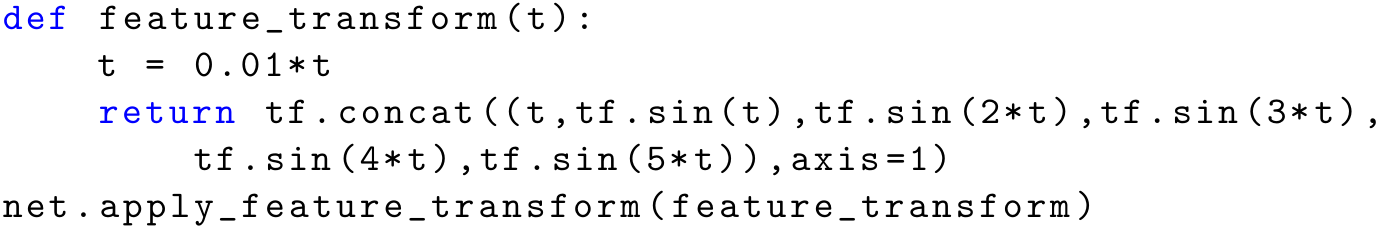}
\caption{\textbf{Input scaling and feature transform.} We use the periodicity of $\sin$ as our feature.}
\label{code:feature_transform}
\end{figure}

\item Output-scaling layer. The outputs $\hat{x}_1, \hat{x}_2, \dots, \hat{x}_S$ may have a disparity of magnitudes. As such, we can scale the network outputs by $\hat{x}_1 = k_1\tilde{x}_1$, $\hat{x}_2 = k_2\tilde{x}_2$, $\dots, \hat{x}_S = k_S\tilde{x}_S$ like in Fig.~\ref{code:output_transform}, where $k_1, k_2, \dots, k_S$ are the magnitudes of the ODE solution $x_1, x_2, \dots, x_S$, respectively.

\begin{figure}[htbp]
\centering
\includegraphics{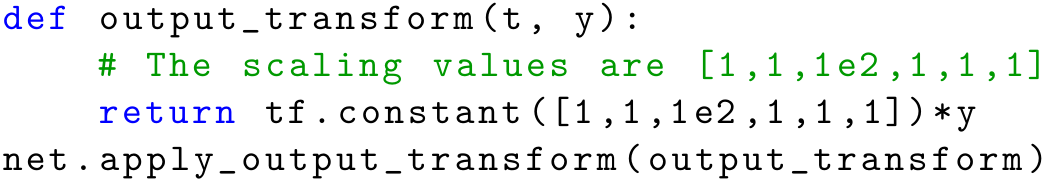}
\caption{\textbf{Output transform to scale the outputs of the network.}}
\label{code:output_transform}
\end{figure}
\end{itemize}

To train the neural network, we need to constrain it to the system of ODEs and the observations we created. This is done by defining a loss function, which computes the difference between the output of the neural network and the desired behavior: following the data at the time $t_1, t_2, \dots, t_{N^{data}}$ and the ODEs at time points $\tau_1, \tau_2, \dots, \tau_{N^{ode}}$. The ODE time points could be chosen at random or uniformly spaced. We define the total loss as a function of $\boldsymbol{\theta}$ and $\mathbf{p}$:
\begin{equation*}
\mathcal{L(\boldsymbol{\theta}, \mathbf{p})} = \mathcal{L}^{data}(\boldsymbol{\theta}) + \mathcal{L}^{ode}(\boldsymbol{\theta}, \mathbf{p}) + \mathcal{L}^{aux}(\boldsymbol{\theta}).
\end{equation*}
$\mathcal{L}^{data}$ is defined for $M$ sets of observations $\mathbf{y}$:
\begin{equation*}
\mathcal{L}^{data}(\boldsymbol{\theta}) = \sum_{m=1}^M w^{data}_m \mathcal{L}^{data}_m = \sum_{m=1}^M w^{data}_m \left[\frac{1}{N^{data}} \sum_{n=1}^{N^{data}} \left(y_m(t_n) - \hat{x}_{s_m}(t_n;\boldsymbol{\theta})\right)^2 \right].
\end{equation*}
$\mathcal{L}^{ode}$ is for our system of ODEs:
\begin{equation*}
\mathcal{L}^{ode}(\boldsymbol{\theta}, \mathbf{p}) = \sum_{s=1}^S w^{ode}_s \mathcal{L}^{ode}_s = \sum_{s=1}^S w^{ode}_s \left[ \frac{1}{N^{ode}} \sum_{n=1}^{N^{ode}} \left( \frac{d\hat{x}_s}{dt} |_{\tau_n} - f_s\left(\hat{x}_s(\tau_n;\boldsymbol{\theta}),\tau_n;\mathbf{p}\right)\right)^2 \right].
\end{equation*}
The last term in the total loss function is $\mathcal{L}^{aux}$, which is used for additional information on system identification. For example, here we assume that we have the measurements of the state variables at two distinct times $T_0$ and $T_1$. While this is essentially a part of the data loss, we include it as its own loss function due to it being given for all state variables at the two time instants. Here, we use the initial condition for $T_0$ and the final time instant for $T_1$, and other choices of $T_0$ and $T_1$ can also be used depending on the available data information.
\begin{equation*}
\mathcal{L}^{aux}(\boldsymbol{\theta}) = \sum_{s=1}^S w^{aux}_s \mathcal{L}^{aux}_s = \sum_{s=1}^S w^{aux}_s \frac{(x_s(T_0) - \hat{x}_s(T_0;\boldsymbol{\theta}))^2 + (x_s(T_1) - \hat{x}_s(T_1;\boldsymbol{\theta}))^2}{2}.
\end{equation*}
The weights $w$ were selected such that all parts of the loss function would be of the same order of magnitude.

\begin{figure}[htbp]
    \centering
    \includegraphics{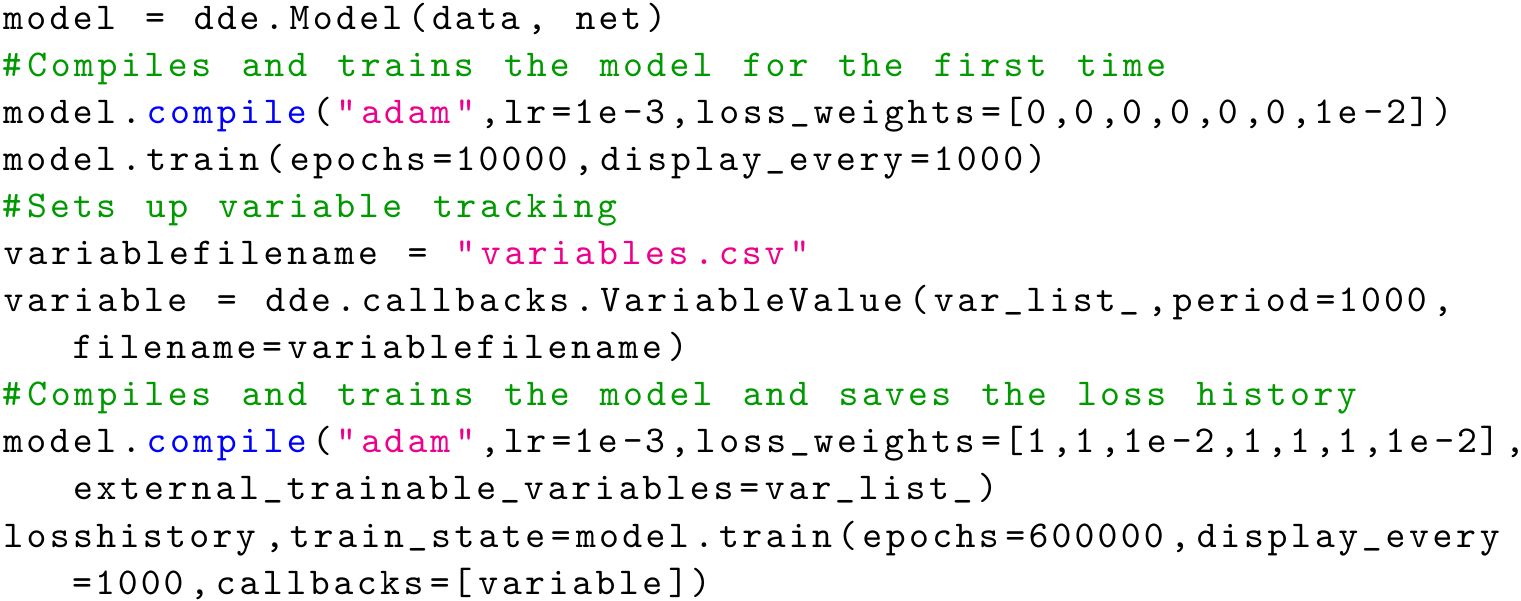}
    \caption{\textbf{Training the model using the Adam optimizer with a learning rate of 0.001.} We train first for 10,000 epochs to let the network understand the data, then for 600,000 for the network to train all the losses.}
    \label{code:train}
\end{figure}

With the loss functions set up, we can train the network and infer the parameters of the ODEs $\mathbf{p}$ by minimizing the loss function via a gradient-based optimizer, e.g., Adam \cite{kingma2014adam}:
\begin{equation*}
\boldsymbol{\theta}^*, \mathbf{p}^* = \arg\min_{\boldsymbol{\theta}, \mathbf{p}} \mathcal{L(\boldsymbol{\theta}, \mathbf{p})}.
\end{equation*}
We first train 10,000 epochs by setting all weights to zero except for data and then train against all parts of the loss function (Fig.~\ref{code:train}). We also track the variables during training by using the callback \texttt{VariableValue}. We also plot the loss function as a function of epochs.

\subsection{Results of SBINN}

\begin{figure}[htbp]
    \centering
    \includegraphics{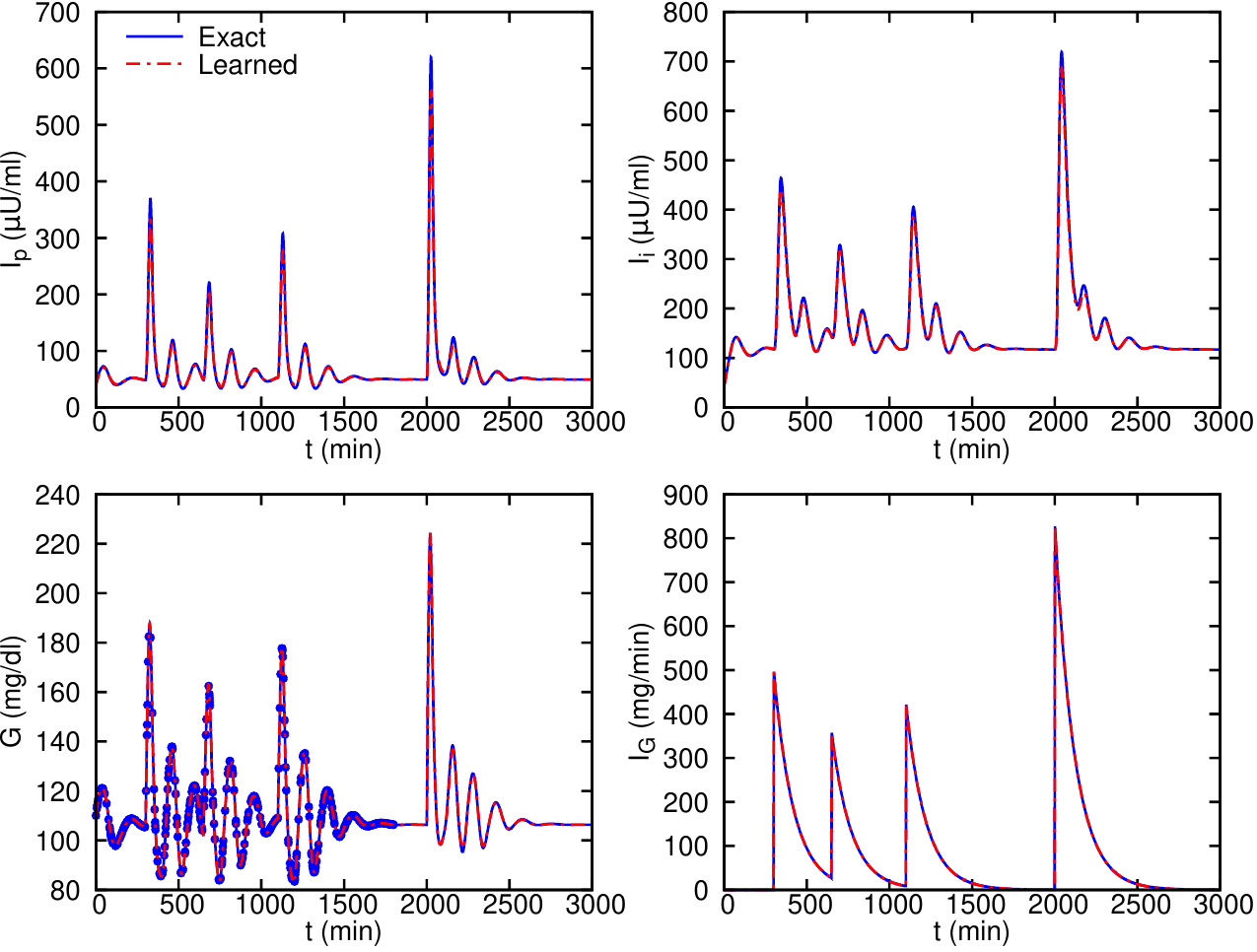}
    \caption{\textbf{Inferred dynamics and forecasting via SBINN.} The network learned the system from time $t = [0,1800]$. The estimated parameters were able to accurately forecast a meal event at $t_j = 2000 \ min$. Figure is adapted with permission from \cite{yazdani2020systems}.}
    \label{fig:glucose-output}
\end{figure}

The inferred parameters are given in Table \ref{table:glucose-app}. We observe good agreement between the inferred values and the target values. We next consider the case of a nutrition event at $t_j = 2000 \ min$ with carbohydrate intake of $m_j = 100 \ g$. We then test how well our model can predict this extra nutrition event using the inferred parameters, the results of which are in Fig.~\ref{fig:glucose-output}. With high accuracy, we determined the glucose levels after the nutrition event.

\section{Practical identifiability analysis}
\label{sec:pract-ident}

The Fisher information matrix (FIM) can be used to develop confidence intervals of parameters as well as determine their practical identifiability, assuming parameters are structurally identifiable, as outlined in Fig.~\ref{fig:flowchart}. The main difference between structural identifiability and practical identifiability lies in the fact that structural identifiability analysis is normally conducted before the fitting of the model and is used to study the uniqueness of parameters assuming noiseless observables. On the other hand, practical identifiability analysis is performed a posteriori and is often used to analyze whether the inferred parameter value will be sensitive to noise in the data. As a consequence, we need both analyses to determine whether the fitting result would be reliable.

\begin{figure}[htbp]
    \centering
    \includegraphics{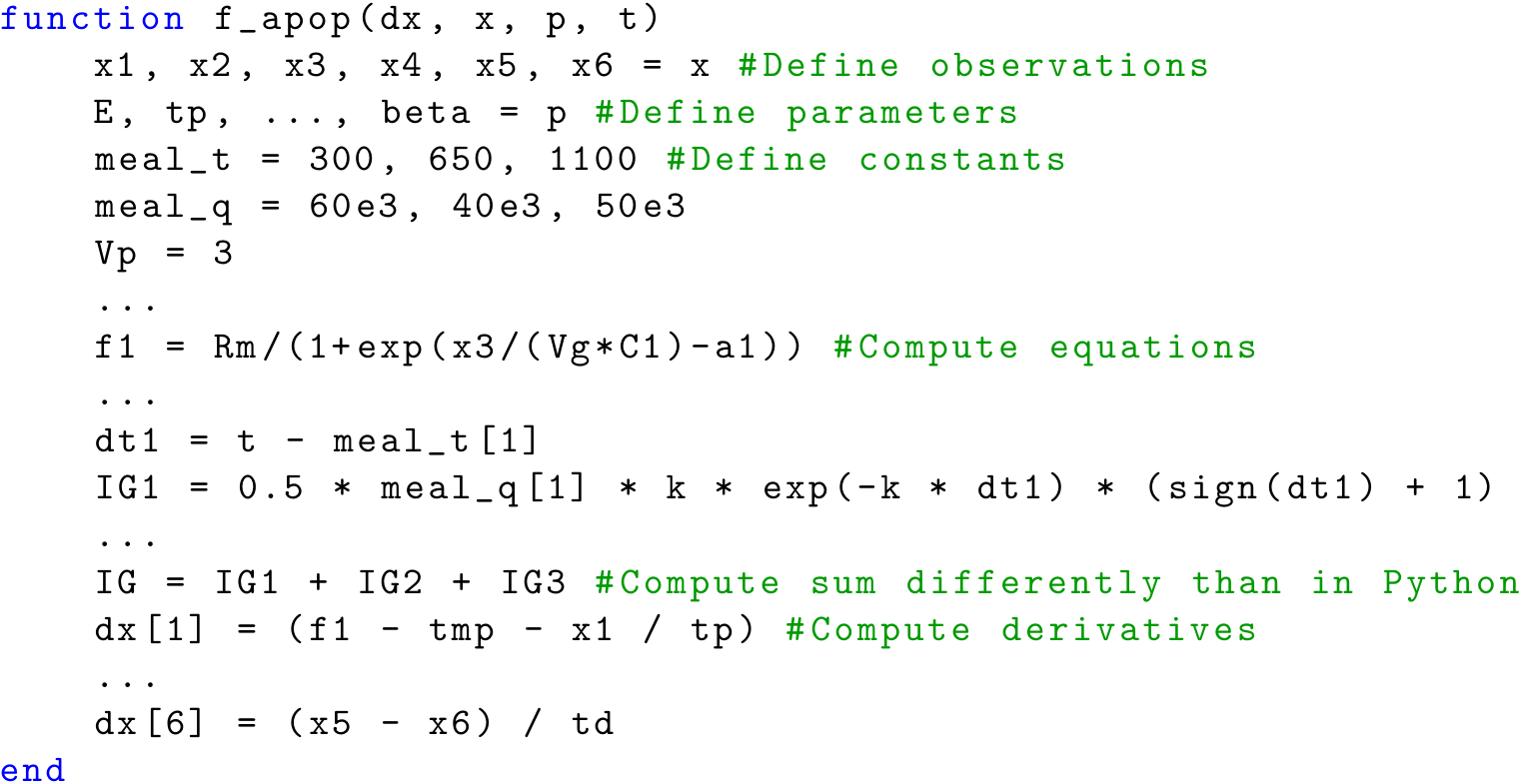}
    \caption{\textbf{Implementation of the ODE system.}}
    \label{code:julia_ode}
\end{figure}

We use Julia for practical identifiability analysis. In Julia, we import the required packages via \texttt{using}. If you are unfamiliar with Julia, see Appendix~\ref{AppendixB}. We start by defining the system in Eqs.~\eqref{eq:glucose1} and \eqref{eq:glucose2}. In our case, $I_p$ is written as \texttt{x1}, and so on.  For their derivatives, we define them as \texttt{dx[1]}, \texttt{dx[2]}, etc. We also declare all parameters that our SBINN determined as a vector \texttt{p}. System definition in Julia is found in Fig.~\ref{code:julia_ode}.

\begin{figure}[htbp]
    \centering
    \includegraphics{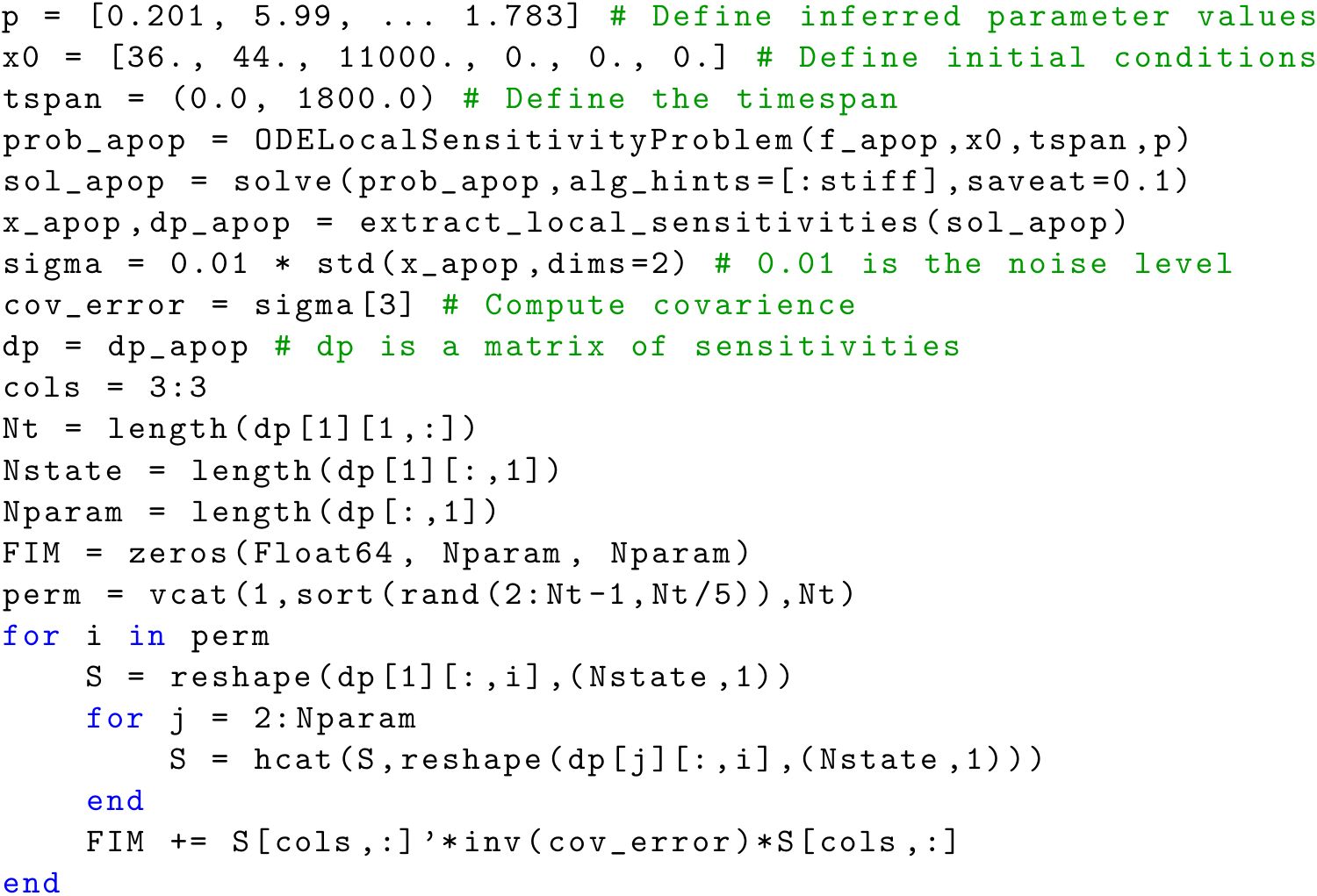} 
    \caption{\textbf{Computing FIM.} \texttt{sigma[3]} in the code refers to the standard deviation of the third state variable $G$.}
    \label{code:FIM}
\end{figure}

To implement the practical sensitivity analysis, we first compute FIM, which is constructed by estimating the sensitivities of the system of ODEs with respect to the parameters. The code to compute FIM is in Fig.~\ref{code:FIM}. We note that even though the data was generated with no noise, we need to consider a noise level of the measurements to compute a meaningful FIM, and we use a low noise level of 1\% in the code. In this example, we only have one observable variable; the code for the problem with more than one observable is almost the same, and we only need to modify \texttt{cov\_error} and \texttt{cols} to indicate the indices of observable variables, see the example and code in \cite{yazdani2020systems}.

There are different ways to utilize FIM, and here we show two important ones: (1) computing the correlation  matrix of all parameters, and (2) computing eigenvectors of FIM associated with the zero eigenvalues (i.e., null eigenvectors).

\begin{figure}[htbp]
    \centering
    \includegraphics[width=12cm]{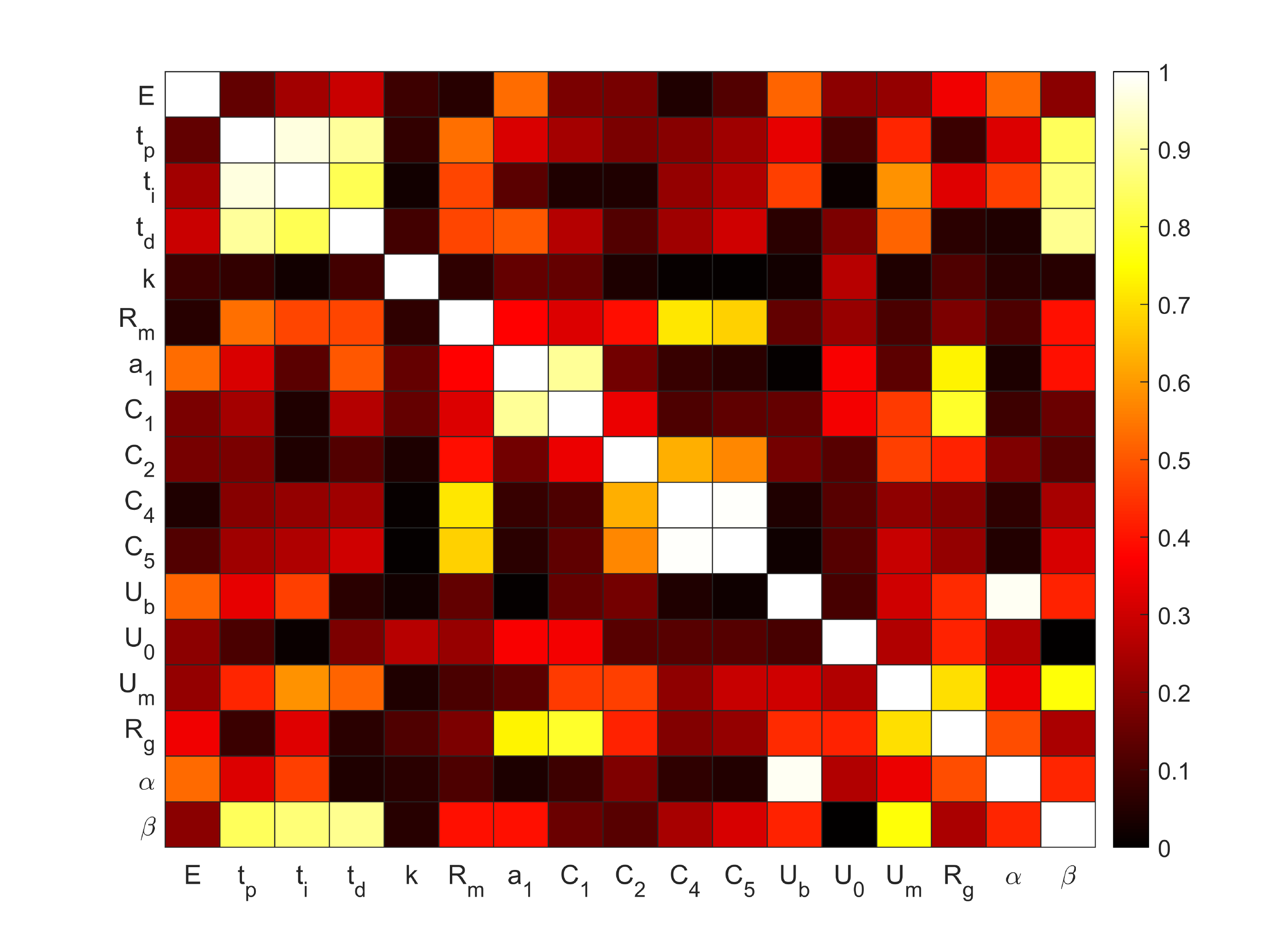}
    \vspace{-10pt}
    \caption{\textbf{Correlation matrix.}} 
    \label{fig:correlation_matrix}
\end{figure}

The correlation matrix $R$ is computed as $R_{ij}$ = $\text{FIM}^{-1}_{ij}/\text{FIM}^{-1}_{ii}$. $|R_{ij}| \approx 1$ indicates that the two parameters $i$ and $j$ are highly correlated, and thus are not individually identifiable from each other. The correlation matrix is shown in Fig.~\ref{fig:correlation_matrix}.

\begin{figure}[htbp]
    \centering
    \includegraphics[width=.9\textwidth]{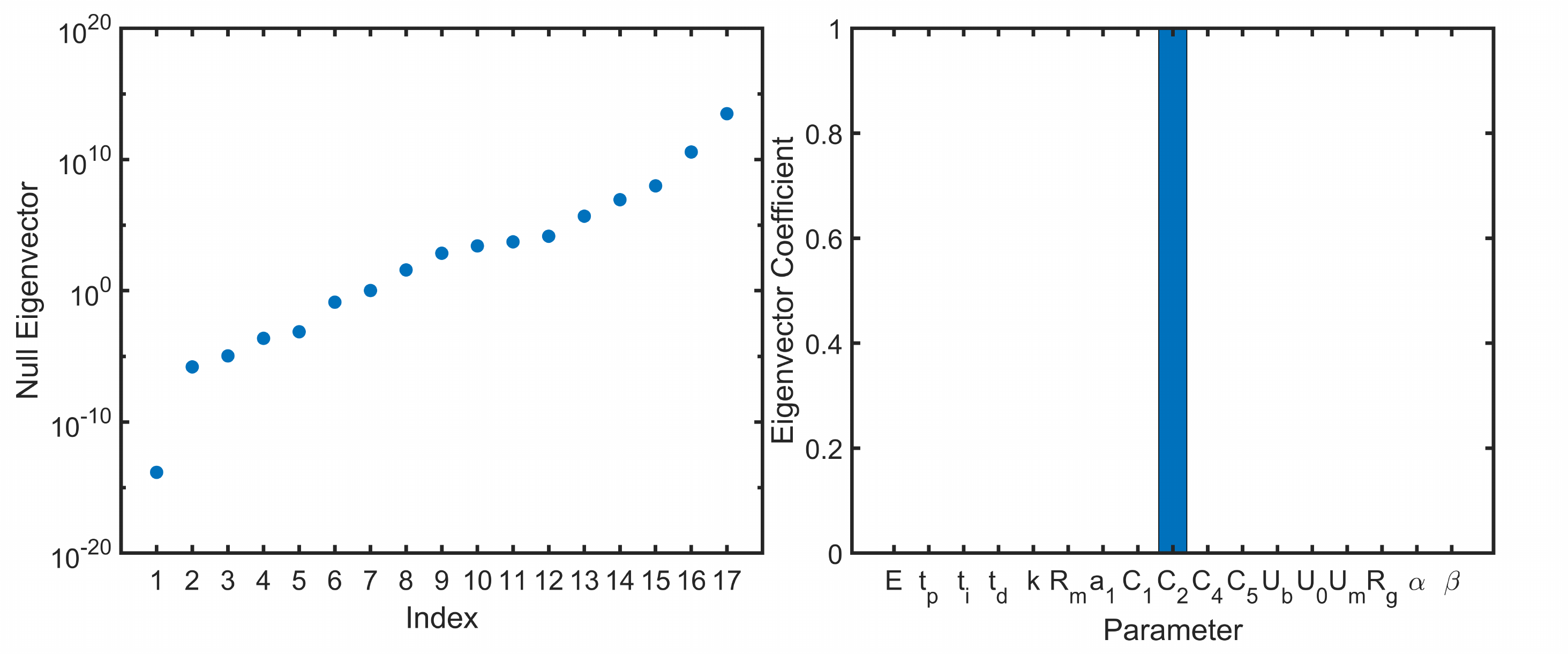}
    \vspace{-15pt}
    \caption{\textbf{Null eigenvector analysis.} (Left) Eigenvalues of FIM. There is one eigenvalue close to 0. (Right) The eigenvector associated with the eigenvalue. The dominant component is $C_2$.}
    \label{fig:eigenvector}
\end{figure}

Next, we compute eigenvalues and eigenvectors of FIM. The eigenvalues is shown in Fig.~\ref{fig:eigenvector} left. There is only one eigenvalue close to 0, and the associated eigenvector (i.e., null eigenvalue) is shown in Fig.~\ref{fig:eigenvector} right, where the value for the $C_2$ component is dominant and all other components are approximately zero. This indicates $C_2$ has little to no effect on state variables and is therefore practically unidentifiable from the dataset.

In this example, the result from the null eigenvector analysis is consistent with our inferred values in Table~\ref{table:glucose-app}, but the correlation matrix is not. We note that FIM-based practical identifiability analysis has many limitations and can be problematic in certain problems. There are other methods available for determining practical identifiability such as the bootstrapping approach \cite{balsacanto2008bootstrap} or using a probabilistic framework to quantify sensitivity of the system dynamics to variations in its parameters \cite{foo2009probabilistic}.

\section{Discussion of time-dependent parameters}

\begin{figure}[htbp]
    \centering
    \includegraphics[width=6cm]{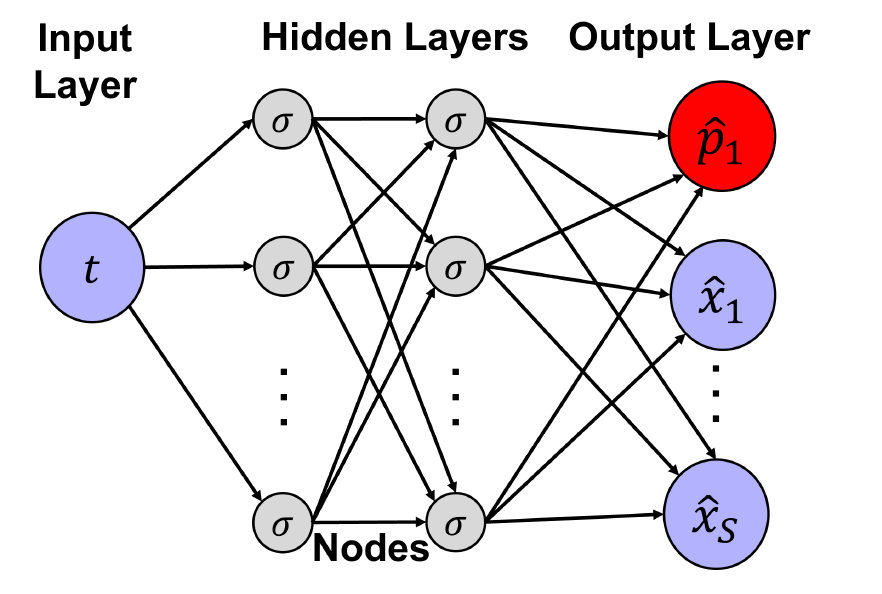}
    \caption{\textbf{SBINN for time dependent parameters.}}
    \label{fig:time_dependent}
\end{figure}

We have considered all parameters are constants, but in real problems, the parameters could vary over time, i.e., time dependent parameters. Here, we briefly describe the idea of implementing time dependent parameters in SBINN. Let us assume $p_1$ is a time dependent parameter to be inferred. We add an extra neuron $\hat{p}_1$ in the network output to represent $p_1$, as shown in Fig~\ref{fig:time_dependent}, and then $\hat{p}_1$ becomes a function of time. Everything else remains the same as the SBINN we introduced in Section \ref{sec:SBINN}.

\section{Summary}

We have provided a complete workflow for analysing biological systems described by a system of ODEs, including structural identifiability analysis, parameter estimation via systems-biology informed neural networks (SBINN), and practical identifiability analysis based on the Fisher information matrix (FIM).

\bibliographystyle{unsrt}
\bibliography{main}

\appendix

\section{Python}
\label{AppendixA}

Python is the most common language for machine learning due to the plethora of libraries available for free. Learning python is fairly easy due to its popularity, and there are a number of free, high quality videos and other tutorials on how to use python.
We note that common software for installing Python is Anaconda, from which most common libraries have already been installed.
The code we provide should remove the majority of the guesswork if solving similar problems to those stated in this tutorial, otherwise, the DeepXDE documentation \url{https://deepxde.readthedocs.io} should be of help.

\section{Julia}
\label{AppendixB}

Julia is another language used for machine learning. It is very similar to Python as far as the syntax is concerned. We recommend using the softwares Atom and the Juno IDE, though Jupiter notebook and similar programs will suffice. There are also a variety of online sources that provide free help for learning this language. If you understand the fundamentals of Python, you should be able to read our provided Julia code and use it for your own situation with only a few minor tweaks that do not involve a heavy amount of coding or even a thorough knowledge of Julia.



\end{document}